\documentclass[
    ,final            
  ]
  {aipproc}

\layoutstyle{6x9}

\begin{document}

\title{Brane-world Cosmology}

\classification{98.80.Cq, 98.80.Jk, 04.50.+h}
\keywords      {Cosmology, extra dimensions, branes, gravity}

\author{David Wands}{
  address={Institute of Cosmology and Gravitation, University of Portsmouth, Mercantile House, Portsmouth P01 2EG, United Kingdom}
  }

\begin{abstract}
 Brane-world models, where observers are restricted to a brane in a higher dimensional spacetime, offer a
 novel perspective on cosmology. I discuss some approaches to cosmology in extra dimensions and some
 interesting aspects of gravity and cosmology in brane-world models.
\end{abstract}

\maketitle


\section{Cosmology after Einstein}

A century after Einstein first proposed his theory of relativity,
it has become a cornerstone of the physical sciences. Four
dimensional spacetime provides the setting for describing physical
processes and in particular provides the dynamical framework for
cosmological models of our expanding Universe.

It was the general theory of relativity, proposed by Einstein in
1915, that for the first time provided equations with which to
describe the dynamics of spacetime. Einstein's equation
\begin{equation}
 \label{EE}
 G_{AB} + \Lambda g_{AB} = 8\pi G_N T_{AB} \,,
\end{equation}
relates the intrinsic curvature, $G_{AB}$, of the metric,
$g_{AB}$, to the local energy-momentum, $T_{AB}$, while allowing
for the possibility of a non-zero cosmological constant,
$\Lambda$. Consistency with Newtonian gravity in the weak-field,
slow-motion limit is ensured by the appearance of Newton's
constant, $G_N$, in the constant of proportionality.

In much of modern cosmology Einstein's tensor equation (\ref{EE})
conveniently reduces to the Friedmann constraint equation
\begin{equation}
 \label{F}
3 \left( H^2 + \frac{K}{a^2} \right) = \Lambda + 8\pi G_N \rho \,,
\end{equation}
which relates the Hubble expansion, $H$, and spatial curvature
$K/a^2$, of a homogeneous and isotropic Friedmann-Robertson-Walker
(FRW) spacetime to the local energy density $\rho$.

Homogeneous and isotropic expansion has been used to build up a
remarkably successful model for the evolution of our Universe
starting with a hot Big Bang at a finite time in our past. This
model has been tested not only by qualitative features such as the
evolution of galaxy populations and the existence of a cosmic
microwave background (CMB) radiation, but also quantitatively
tested by comparing models of primordial nucleosynthesis with
abundance of light elements.

One only needs to consider linear perturbations about a
homogeneous and isotropic metric to build up a coherent picture of
the formation of structure in our Universe. Small fluctuations,
about one part in a hundred thousand, are observed in the
temperature of the microwave background radiation and indicate the
existence of small primordial perturbations in the distribution of
matter and radiation in the early universe when the CMB last
scattered, about 300,000 years after the Big Bang. These
primordial density fluctuations provide the seeds around which the
observed large-scale structure of our Universe can form simply by
gravitational instability, in a cosmological model with
appropriate contributions from radiation, baryonic matter, as well
as cold dark matter and some form of dark energy, that behaves
very much like Einstein's cosmological constant today. A wealth of
observational data now enables cosmologists to put this basic
picture to the test and attempt to measure parameters such as the
density of different forms of matter, the nature of the primordial
perturbations, and Einstein's gravitational laws.

At the same time fundamental questions remain unanswered. Why are
there 3 large spatial dimensions (not 5 or 15)? why is the value
of the cosmological constant so small? and what really happens at
the initial Big Bang which represents a singular point at the
start of our cosmological evolution?
I cannot answer these questions in this talk, but I can show how
brane-world models offer some novel and interesting perspectives
on these issues. 
I should emphasize that this is a personal view and not intended 
to be a systematic review of all aspects of brane-world cosmology.
For a more comprehensive review see~\cite{Roy}.

\section{Extra dimensions}

Superstring theory is an attempt to unify gravity with the other
fundamental interactions in a self-consistent quantum theory,
based on strings (extended 1-dimensional objects) as the
fundamental constituents of matter rather than point particles. In
particular string theory should be finite and singularity free.

For example, the existence of a minimal length scale in the
effective theory leads to a ``T-duality'' that relates expanding
and contracting cosmological solutions and has been proposed as
the basis for the pre-Big Bang scenario \cite{GasVen} that
proposes a pre-Big Bang era preceding the hot Big Bang expansion.
Unfortunately the nature of the transition from pre- to
post-Big Bang is dependent on the nature higher-order, possibly
non-perturbative, effects and remains elusive. This makes it hard
to make robust predictions based on a pre-Big Bang phase.

It is not fair to say that string theory does not make any
predictions.
String theory does make a definite prediction for the number of
spacetime dimensions. Spacetime should have 10 dimensions for a
consistent, anomaly-free superstring theory \cite{GSW}. This may
not appear to be a huge success for the theory, but of we can only
assert that there are four {\em observable} dimensions and it
is quite possible that there exist {\em extra} dimensions that are
very small and/or unobservable.

Only a few years after Einstein proposed his theory of dynamical
four-dimensional spacetime, Kaluza began to consider the dynamical
equations for a five-dimensional spacetime, realising that the
degrees of freedom of the metric associated with the extra
dimension could describe a vector field in our four-dimensional
world \cite{Kaluza}. If the extra dimension is compact and very
small, less than $10^{-19}$~m say, then only the zero-mode of
the metric, or other fields, would be excited in terrestrial
experiments. Higher harmonics in hidden dimension(s) correspond to
very massive states, requiring large energies to excite them, and
these can be consistently set to zero in a low-energy effective action.

In time it was realised that the size of the extra dimension was
itself a scalar field and higher-dimensional models of gravity
reduce to an effective scalar-tensor theory of gravity in
four-dimensions at low energies. To avoid conflict with
experimental tests of gravity the size of the extra dimensions
must be fixed, but there has been little progress on how to
stablise all the moduli fields describing the size and shape of
hidden dimensions in string theory, until recently.

This ``hosepipe view'' of the extra dimensions being rolled up
incredibly small and hence out of sight was almost universally
adopted to deal with the embarrassment of extra dimensions in
string theory until the 1990's.
What changed in the mid 1990's was that realisation that other
extended objects, higher-dimensional membranes, or ``branes'',
should also play a fundamental role in string theory
\cite{Polchinski}. Branes opened up the possibility to related
apparently different string theories, for instance string theories
containing closed strings or those with open strings.

Branes can support open strings whose end-points lie on the brane.
These open strings can describe matter fields which live on the
brane. On the other hand perturbations of the higher dimensional
bulk geometry are described by excitations of closed strings,
such as the graviton. To a general relativist it should be clear
that even if matter fields are restricted to a lower dimensional
hypersurface, gravity as a dynamical theory of geometry must exist
throughout the spacetime.


\begin{figure}
  \includegraphics[height=.3\textheight]{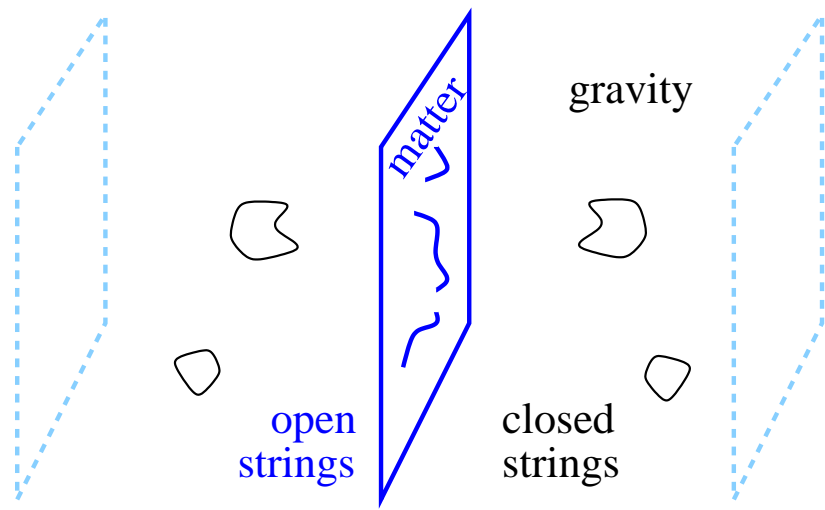}
  \caption{Matter fields are described by open strings confined to branes.}
\end{figure}


This lead several authors to consider the possibility that at
least some of the extra dimensions could be far larger than had
previously been imagined \cite{A,AADD}. They realised that while
particle interactions are probed by high energy colliders on
energies up to $1$~TeV, and hence scales down to $10^{-19}$~m,
gravity is barely tested on scales below $1$~mm.
If the extra dimensions were testable only via gravity then they
might be relatively large. This offers a tantalising explanation
for why gravity appears to be so weak when compared with the other
interactions. The gravitational field of an object could leak out
into the large but hidden dimensions and gravity in our
four-dimensional world seems weaker.

To make this a little more precise, consider the gravitational
field of a mass $M$ in a $D$-dimensional spacetime. If we use
Gauss's law to calculate the gravitational field strength $g$ at a
distance $r$ then we find $g\propto G_DM/r^{D-2}$ for distances
$r\ll R$, the radius of compactification of the hidden dimensions.
But if $r\gg R$ then the gravitational field strength is given by
\begin{equation}
 g = \frac{4\pi G_DM}{4\pi r^2 R^{D-4}} \,.
 \end{equation}
The effective value of Newton's constant in our apparently
4-dimensional world, $G_4$, can be identified as
\begin{equation}
 G_4 \equiv \frac{G_D}{R^{D-4}} \,.
\end{equation}

Given we observe only the four-dimensional effective gravitational
coupling, from which we infer a very large effective Planck scale
$M_4=1/\sqrt{G_4}=10^{19}$~GeV, the true value of the Planck scale
(the scale at which quantum gravity becomes important) could be
much smaller in models with large extra dimensions.

For instance, Horava and Witten in 1996 \cite{HW96} proposed a
supergravity model in 11-dimensions with a fundamental Planck
scale close to the Grand Unified (GUT) scale of $10^{16}$~GeV
where one of the extra dimensions had a size
considerably larger than the conventional Planck scale of
$10^{-35}$~m.
But the GUT scale is still far beyond terrestrial experiments and
established particle physics models. What if quantum gravity was
within reach of experiments like the LHC at CERN? If one hidden
dimension was as large as $1$~mm then the Planck scale could be as
low as $10^8$~GeV. With two large extra dimensions, the Planck
scale could be as low as $1$~TeV \cite{AADD}.

\section{Randall-Sundrum model}

So far I have implicitly been discussing Minkowski branes in
a higher dimensional Minkowski spacetime. This provides a good
vacuum state for string theory but we need to go beyond flat
spacetime to provide a cosmological model. Anti-de Sitter (AdS)
spacetime, that is maximally symmetric space with a negative
cosmological constant, $\Lambda^2=-6k^2$, can also provide a
useful vacuum state for string theory. This may not appear to be
very promising for a cosmological model as a negative cosmological
constant leads to a cosmological collapse and big crunch in
homogeneous and isotropic cosmologies. However it turns out to be
a fascinating spacetime in which to consider brane-world
cosmology.

Randall and Sundrum produced two papers \cite{RS1,RS2} in 1999
which have had a huge impact in string theory and cosmology. They
considered gravity on constant tension branes embedded in
five-dimensional anti-de Sitter spacetime.

Branes can be embedded at fixed $y$-coordinate in a Gaussian
normal coordinate system where the AdS$_5$ metric is written as
\begin{equation}
 \label{ads5}
ds^2 = e^{-2k|y|} \eta_{\mu\nu} dx^\mu dx^\nu + dy^2 \,.
\end{equation}
The exponential ``warp factor'' means that the volume of the extra
dimensional space becomes small at large $y$. In their first paper
\cite{RS1} Randall and Sundrum showed that the large hierarchy
between a fundamental TeV scale and the apparent Planck scale
$10^{19}$~GeV could be ``explained'' by a large warp factor even
if the size of the extra dimension (specifically the normal
distance between branes) was relatively small. But in their second
paper \cite{RS2} they showed that even if there was no second
brane, and the extra dimension extended to infinity, gravity
remained effectively localised on a single brane as the integrated
volume remained finite as $y\to\infty$. This they proposed as an
``alternative to compactification''.


\begin{figure}
  \includegraphics[height=.3\textheight]{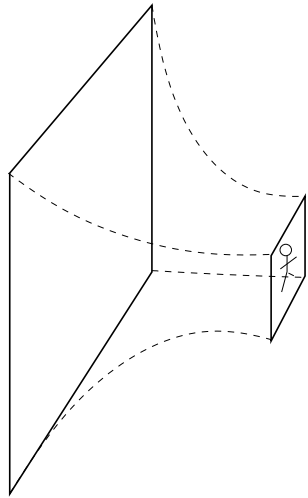}
  \caption{Randall Sundrum 1, where the hierarchy between the Planck scale and the TeV scale is due to the distance between two branes in compact AdS spacetime.}
\end{figure}



\begin{figure}
  \includegraphics[height=.3\textheight]{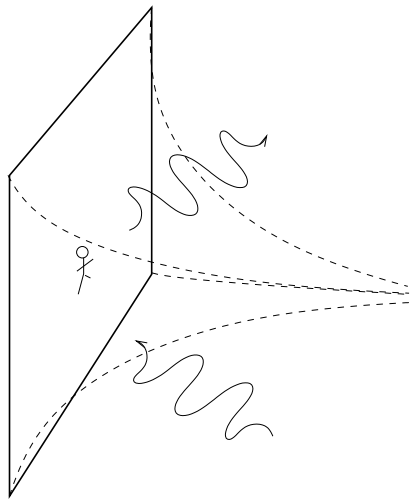}
  \caption{Randall Sundrum 2, where there is only one brane embedded in non-compact 5D spacetime.}
\end{figure}


The two-brane model \cite{RS1}, called RS1, is not so different
from earlier attempts to compactify the hidden dimensions, other
than that it operates in a curved bulk spacetime. It is still the
large volume of the hidden space that makes gravity weaker on the
brane than other forces. There is still a discrete spectrum of
Kaluza-Klein states corresponding to higher harmonics on the
hidden space, although the spectrum of eigenvalues is different in
a curved space. And the size of the extra dimension, the distance
between the two branes remains a scalar degree of freedom, known
as the radion. It still leads to an effective scalar-tensor
gravity in four dimensions at low energies \cite{GT,KS} which may
be in conflict with experimental tests unless the radion is
stabilised.

On the other hand, the one-brane model \cite{RS2}, inevitably
known as RS2, offers a radically different model of dimensional
reduction. The radion field in the RS1 model, decouples from
gravity on the remaining brane in the limit that the second brane
tends to spatial infinity. (The Brans-Dicke parameter
$\omega\to\infty$ \cite{GT,KS}.) And the discrete spectrum of KK
modes is replaced with a continuum of bulk modes. However the
lightest modes are only weakly coupled to matter on the brane and
gravity remains effectively four-dimensional on length scales
greater than the AdS curvature scale, $k^{-1}$. More fundamentally
though the single brane in AdS is an open system now where the
initial state of matter on the brane (or branes) is not enough to
determine the future evolution of the system. Instead one needs to
specify initial data on a Cauchy hypersurface in the bulk. For
example one might specify the AdS incoming vacuum state
\cite{Rubakov}. And fields on the brane can radiate into the bulk
and information can escape to future null infinity.

In either of the RS models there is a simple and novel
interpretation of our cosmological expansion. In the curved
anti-de Sitter bulk spacetime (\ref{ads5}) any motion of the
brane, represented by a time-dependent trajectory $y=y_b(t)$,
induces an FRW metric on the brane with scale factor $a=e^{-k|y_b|}$
\cite{MSM,BCG}. Cosmological expansion on the brane corresponds to
motion in a curved bulk spacetime.

\section{Brane-world gravity}

One way to understand the gravitational theory on a brane,
such as the Randall-Sundrum branes in AdS, is to use the projected
Einstein equations on the brane \cite{SMS,Roy}.
Consider a codimension-one brane with unit normal vector $n^A$.
The induced metric on the brane is then
\begin{equation}
 g_{AB} = \, ^{(5)}g_{AB} - n_A n_B \,,
\end{equation}
and the extrinsic curvature of the brane is
\begin{equation}
 K_{AB} = g^{AC}\, ^{(5)}\nabla_C n_B \,.
\end{equation}
The 4D Riemann tensor on the brane can be given in terms of the 5D
Riemann tensor in the bulk and the brane's extrinsic curvature as
\cite{Roy}
\begin{equation}
 \label{Riemann}
R_{ABCD} =  ^{(5)}R_{EFGH} \, g_A^E \, g_B^F \, g_C^G \, g_D^H + 2
K_{A[C}K_{D]B} \,.
\end{equation}

The higher-dimensional Einstein equations (\ref{EE}) determine the
bulk Einstein tensor in terms of the bulk energy-momentum tensor.
In the case of a vacuum bulk with only a cosmological constant we
have
\begin{equation}
^{(5)}G_{AB} + \Lambda_5 \, ^{(5)}g_{AB} = 0 \,.
\end{equation}
The Israel-Darmois junction conditions determine the jump in the
extrinsic curvature tensor across the brane in terms of the
energy-momentum tensor localised on the brane where $\kappa_5^2$
is the gravitational coupling constant in 5-dimensions.
In the Randall-Sundrum model the brane is a boundary of the bulk
spacetime. This is equivalent to imposing a $Z_2$-symmetry across
the brane so that $K_{AB}=K_{AB}^+=-K_{AB}^-$ and hence
\begin{equation}
 \label{KAB}
 K_{AB}
  = - \frac{\kappa_5^2}{2}
   \left[ T^{\rm brane}_{AB} - \frac13 T^{\rm brane} g_{AB} \right] \,.
\end{equation}
This also occurs in the Horava-Witten model where the 10D boundary
branes are fixed points of the orbifold $S_1/Z_2$. On the other
hand the HW model also admits additional branes which can move
within the bulk spacetime. In this case there is an additional
freedom due to the averaged extrinsic curvature,
$K_{AB}^++K_{AB}^-$, which is not directly constrained by the
energy-momentum tensor on the brane \cite{AndyMetal}, but for
simplicity I will assume $Z_2$-symmetry across the brane in the
following.

Finally putting all this together we can give an expression for
the Einstein tensor for the induced metric on the brane \cite{SMS}
\begin{equation}
 \label{braneEE}
 G_{AB} + \Lambda g_{AB} = \kappa_4^2 T_{AB} + \kappa_5^4 {\cal S}_{AB} -
 {\cal E}_{AB} \,,
\end{equation}
where (i) ${\cal S}_{AB}$ and (ii) ${\cal E}_{AB}$ represent
modifications to the standard Einstein equations (\ref{EE}) due to
(i) terms quadratic in the brane energy-momentum tensor and (ii)
the 5D Weyl tensor projected on the brane.

The 4D intrinsic curvature (\ref{Riemann}) includes terms
quadratic in the extrinsic curvature of the brane, and hence, via
(\ref{KAB}), the energy-momentum on the brane. Indeed we only
recover a term linear in $T_{AB}$ in Eq.~(\ref{braneEE}) if the
energy-momentum tensor on the brane contains a constant part due
to a constant tension or vacuum energy density on the brane,
$\sigma$, so that we split
\begin{equation}
T_{AB}^{\rm brane} = \sigma g_{AB} + T_{AB} \,.
\end{equation}
The effective 4D gravitational coupling constant for the
renormalised energy-momentum tensor, $T_{AB}$, in the brane-world
Einstein equations (\ref{braneEE}) is then given by
\begin{equation}
\kappa_4^2 = \frac{\kappa_5^4\sigma}{6} \,.
\end{equation}

The effect of terms quadratic in the matter energy-momentum tensor
is given by
\begin{equation}
 {\cal S}_{AB} = \frac{1}{12} TT_{AB} - \frac14 T_{AC}T_B^C +
 \frac{1}{24} g_{AB} \left( 3T_{CD}T^{CD}-T^2 \right) \,.
\end{equation}
It is represents a high-energy correction to the brane-world
Einstein equations and is typically unimportant when the matter
density is much less than the brane tension, $\rho\ll \sigma$.

\subsection{The brane-world cosmological constant problem}

The effective cosmological constant on the brane in
Eq.~(\ref{braneEE}) is given by
\begin{equation}
\Lambda = \frac{\Lambda_5}{2} + \frac{\kappa_5^4\sigma^2}{12} \,.
\end{equation}
In contrast to our usual 4D viewpoint that the vacuum energy
density should simply vanish, or be very small, in the brane-world
we require instead that there is a cancellation between the 4D and
5D contributions to the vacuum energy.
An intriguing possibility in the brane-world is that 4D
cosmological solutions might naturally seek out fixed points in
an inhomogeneous 5D spacetime with small values of the cosmological constant --
called self-tuning solutions \cite{selftune}.

A novel twist on the cosmological constant problem is provided by
the model of Dvali, Gabadadze and Porrati (DGP) \cite{DGP} who
pointed out that quantum loop corrections to any classical model
would be expected to induce terms in the effective energy-momentum
tensor on the brane proportional to the brane Einstein tensor:
$\Delta T_{AB}^{\rm brane}= \alpha G_{AB}$. In a 4D model such
corrections would simply renormalise the gravitational coupling
$\kappa_4^2$. But substituted into (\ref{braneEE}) the brane-world
Einstein equations become quadratic in the Einstein tensor. Thus
in addition to the usual vacuum solution with $G_{AB}=0$ when
$\Lambda=0$, there is a second (non-perturbative) solution with
$G_{AB}\propto \alpha^{-2}g_{AB}$. The DGP model has sparked great
interest as a novel explanation of the observed acceleration of
our Universe \cite{Deffayet}, in terms of modified gravity rather
than ``dark energy'', but there remain questions over whether
the self-accelerating solutions admit unstable ``ghost''
modes \cite{KK}.

\subsection{Non-local brane gravity}

Equation~(\ref{braneEE}) leaves only the projected 5D Weyl tensor
\begin{equation}
{\cal E}_{AB} = \, ^{(5)}C_{ECFD} \, g_A^E \, g_B^F \, n^C \, n^D
\,.
\end{equation}
undetermined by the local energy-momentum on or near the brane.
This is the tidal part of the 5D gravitational field so is only
determined when one has a solution to the full 5D Einstein
equations with appropriate boundary conditions.

To the brane-bound observer it may be interpreted as an effective
``Weyl fluid'' with energy density $\tilde\rho$ and 4-velocity
$\tilde{u}_A$ so that \cite{Roy}
\begin{equation}
- {\cal E}_{AB} = \kappa_4^2 \left[ \frac{\tilde\rho}{3}
 \left( g_{AB} + 4 \tilde{u}_A \tilde{u}_B \right) +
 \tilde\Pi_{AB} \right] \,.
\end{equation}
Because of the symmetries of the bulk Weyl tensor, ${\cal E}_{AB}$
is trace-free and hence the Weyl fluid is trace-free and has be
interpreted as ``dark radiation'' \cite{Kraus}. This is consistent
with the Maldacena's AdS-CFT conjecture \cite{Maldacena} which
implies that the higher-dimensional gravitational field is
equivalent to a conformal field theory on the boundary
\cite{Gubser}.

The 4D Bianchi identities, $\nabla^AG_{AB}=0$, imply from
Eq.~(\ref{braneEE}) that the Weyl fluid's energy $\tilde\rho$ and
momentum $\tilde\rho\tilde{u}_A$ obey local conservation equations
on the brane, driven by the quadratic energy-momentum tensor
\begin{equation}
 \nabla^A {\cal E}_{AB} = \kappa_5^4 \nabla^A {\cal S}_{AB} \,.
 \end{equation}
The evolution of the Weyl anisotropic stress, $\tilde\Pi_{AB}$,
however cannot in general be determined from initial conditions
set solely on the brane. Thus while the projected equations, and
the Weyl fluid description in particular, may be useful for
interpreting 5D gravity as seen on the brane, it may be of limited
use in deriving solutions. This intrinsic non-locality of 4D
gravity in the brane-world is the reason why so many outstanding
problems remain, including the nature of black hole solutions on
the brane or anisotropic cosmologies, and require
higher-dimensional solutions.

\section{FRW cosmology on a brane}

One, important, case in which the projected field equations are
sufficient is the behaviour of 4D homogeneous and isotropic (FRW)
cosmologies. In this case the maximal symmetry of 3D space
requires that the Weyl anisotropic stress vanishes and the general
form of the modified Friedmann equation (\ref{F}) on the brane is
\begin{equation}
 \label{braneF}
3 \left( H^2 + \frac{K}{a^2} \right) = \Lambda + \kappa_4^2 \rho
\left( 1 + \frac{\rho}{2\sigma} \right) + \frac{m}{a^4} \,,
\end{equation}
where $m$ is an integration constant on the brane set by the
initial density of the Weyl fluid or dark radiation on the brane.

In fact a generalisation of Birkhoff theorem implies that the
general 5D vacuum spacetime admitting an FRW brane cosmology is
Schwarzschild-Anti-de Sitter (SAdS) \cite{MSM,BCG}. The
integration constant $m$ in (\ref{braneF}) represents the mass of
the black hole in the SAdS spacetime (though in a compact RS1
model the singularity may lie outside the physical region of the
spacetime between the branes).

The cosmological expansion described by the modified Friedmann
equation (\ref{braneF}) can be interpreted as motion of the brane
in a static, but curved bulk. The static bulk metric can be
written as
\begin{equation}
 \label{dsTR}
ds^2 = -f(R)dT^2 + \frac{dR^2}{f(R)} + R^2 d\Omega_K^2 \,,
\end{equation}
where $d\Omega_K^2$ is the line element on a maximally symmetric
3-space, curvature $K$, and
\begin{equation}
f(R) = K + \left( \frac{R}{\ell} \right)^2 - \frac{m}{R^2} \,.
\end{equation}
On the other hand if we choose a Gaussian normal coordinate in
which the brane is at a fixed location $y=y_b$ then the line
element becomes
\begin{equation}
 \label{dstauchi}
ds^2 = -n^2(\tau,\chi)d\tau^2 + d\chi^2 + a^2(\tau,\chi)
d\Omega_K^2 \,,
\end{equation}
where the explicit forms of $n$ and $a$ are given in
Ref.~\cite{BDEL}.
The two coordinate systems are related by a pseudo-Lorentz
transformation at the brane \cite{LMW}
\begin{equation}
\left( \begin{array}{c} n d\tau \\ d\chi \end{array} \right) =
\Lambda(\theta)
 \left( \begin{array}{c} \sqrt{f}dT \\ dR/\sqrt{f} \end{array} \right)
 \,,
\end{equation}
where
\begin{equation}
 \label{Lorentz}
\Lambda(\theta) \equiv
 \left( \begin{array}{cc} \cosh\theta & \sinh\theta
  \\ \sinh\theta & \cosh\theta \end{array} \right)
\end{equation}
and the Lorentz factor due to the motion of the brane in the bulk
coordinates is
\begin{equation}
\cosh\theta = \sqrt{1+ \frac{R^2H^2}{f}} \,,
\end{equation}
and $H$ is the Hubble expansion rate (\ref{braneF}).

This offers a novel perspective on 4D cosmology, not least the
cosmological singularity problem. For instance, Garriga and Sasaki
showed that an inflating brane and its SAdS bulk can be created
``out of nothing'' by a de Sitter-brane instanton \cite{GarrigaSasaki}. Others have
tried to describe the big bang singularity on the brane as a
singular event within a regular higher-dimensional bulk, see for
example Refs.~\cite{ekpyrotic,pyro,cyclic,bucher,bbbbbb}.

\subsection{Colliding FRW branes}

One scenario that has attracted much attention is the ekpyrotic
model \cite{ekpyrotic,pyro} where the Big Bang on the brane-world
is identified as a collision between branes. In the original
version a hidden brane traverses the bulk and when it hits the
boundary brane its kinetic energy is released, heating our
observable universe and initiating the hot Big Bang.

\begin{figure}
 \label{3branes}
  \includegraphics[height=.3\textheight]{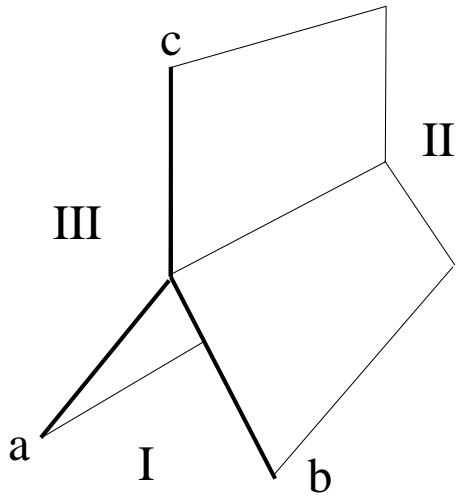}
  \caption{Collision between branes (a) and (b), separated by region I, resulting in single outgoing brane (c) between regions II and III.}
\end{figure}

It turns out to be possible to give a complete description in
general relativity of the collision of maximally symmetric
codimension-one branes (or shells) in vacuum \cite{collide}, 
similar to that envisaged in the original ekpyrotic model. Consider the
simplest case of two incoming FRW brane-worlds (a and b)
coalescing at a 3D collision surface to give one outgoing brane
(c) as shown in Figure~\ref{3branes}. The intervening regions (I,
II and III) are necessarily SAdS in vacua. Thus the coordinate
system on each brane (\ref{dsTR}) and in each bulk region
(\ref{dstauchi}) are related by pseudo-Lorentz transformations of
the form given in (\ref{Lorentz}).


There is a simple geometrical constraint that the product of all
the Lorentz transformations (\ref{Lorentz}), as one completes a
circuit around the collision surface, is unity \cite{Neronov,collide}
\begin{equation}
 \label{product}
\Pi_i \Lambda(\theta_i) = 1 \,.
\end{equation}
The junction conditions (\ref{KAB}) enable one to relate the
corresponding jump in the extrinsic curvature across each brane to
the energy density on the brane, and hence one obtains a general
relativistic version of the local conservation of energy-momentum
at the collision \cite{collide}
\begin{equation}
 \label{energycons}
\rho_c \cosh\theta_c = \rho_a\cosh\theta_a + \rho_b\cosh\theta_b
 \,.
\end{equation}
It is remarkable that a geometrical identity (\ref{product}) when
combined with the general relativistic junction conditions
(\ref{KAB}) reduces to a formula (\ref{energycons}) that is of
exactly the same form as energy conservation in special
relativity.

There is no curvature singularity at the brane collision and yet
the collision does represent an abrupt change in the energy
density and expansion rate on the brane.

In the original ekpyrotic scenario \cite{ekpyrotic} the collision
occurs between a moving bulk brane and a $Z_2$-symmetric boundary
brane. The $Z_2$-symmetry at the boundary makes this equivalent to
two symmetric bulk branes hitting the boundary at the same
collision point. In principle one should also be able to calculate
the spectrum of primordial perturbations inherited after the
collision and thus test the model, see for instance
Refs.\cite{ekpyrotic,lyth}. However, in the subsequent cyclic
scenario \cite{cyclic} the collision takes place between two
$Z_2$-symmetric boundary branes, which is equivalent to an
infinite number of branes colliding at the same hypersurface with
an infinite energy density. Equivalently the fifth dimension
disappears at the collision. This means that the collision really
is singular even from the five-dimensional viewpoint and one
cannot calculate the properties of any brane-world that emerges
from the collision in any conventional (e.g., general
relativistic) theory. Instead one has to construct a non-singular
``completion'' of the theory in order to eliminate the diverges,
see, for example Ref.\cite{tolley}.

\subsection{Brane inflation}

Branes have had a big impact in recent years upon attempts to
construct models of inflation in the early universe within string
theory. A long-standing problem has been the large number of light
scalar fields, or moduli in the low-energy effective actions
obtained from string theory. These would not only violate
experimental tests of gravity today, as mentioned earlier, but also
dilute the vacuum energy density required to drive inflation in
the very early universe. But in recent years the inclusion of
non-trivial anti-symmetric form fields on the hidden dimensions,
so-called flux compactifications \cite{GKP}, offer the possibility of
stabilising all the moduli associated with the shape of the extra
dimensions.
The existence of branes plays a crucial role in providing a
positive vacuum energy density in the low-energy effective theory.

Some of the resulting higher-dimensional geometries bear a close
resemblance to the simple Randall-Sundrum brane-worlds.
Near the branes the ten-dimensional spacetime may be distorted
into a ``throat'' that is approximately described by
AdS$_5\times$S$_5$. This provides the setting, in the KKLT
scenario \cite{KKLT}, for brane-inflation \cite{DvaliTye} in which
the motion of the test branes in this warped geometry plays the
role of an inflaton field. The collision of the branes signals the
end of inflation and can also lead to a symmetry breaking phase
transition on the brane.

These brane inflation models are constructed from low-energy
effective potentials in a four-dimensional effective action. The
dynamics during inflation is similar to models of hybrid inflation
in four dimensional Einstein gravity originally derived from
supergravity models in the 1990's \cite{hybrid}. In particular the
moving branes are test branes moving in a fixed background
geometry, quite different from the self-gravitating branes in
higher-dimensional gravity that I discussed earlier. It remains to
be seen if it is possible to describe popular models of brane
inflation in the covariant approach of Shiromizu, Maeda and
Sasaki, and whether the gravitational backreaction of the branes
can be taken into account, and whether it plays an important role.
For tentative steps in this direction see~\cite{KKKK,Mukohyama,KSW}.

\subsection{Slow-roll inflaton on a brane}

A simple way to study the effect of higher-dimensional gravity on
inflation models is to consider inflation driven by the potential
energy, $V$, of a four-dimensional inflaton field, $\phi$,
confined to a brane, embedded in a vacuum bulk \cite{MWBH}.

At low energies, $\rho\ll \sigma$, we have seen that we can recover the
conventional Friedmann equation for the cosmological expansion. On
the other hand, at high energies the additional quadratic term in
the modified Friedmann equation (\ref{braneF}) leads to
additional Hubble damping and actually assists slow-roll
inflation. The conventional slow-roll parameters, which must be
much smaller than unity for slow-roll inflation, become
\begin{eqnarray}
\epsilon = \frac{1}{16\pi G_N} \left( \frac{V'}{V} \right)^2
 &\to&  \left( \frac{4\sigma}{V} \right) 
 \frac{1}{16\pi G_N} \left( \frac{V'}{V} \right)^2
\,,\\
 \eta = \frac{1}{8\pi G_N} \left( \frac{V''}{V} \right)
 &\to& \left( \frac{2\sigma}{V} \right)
 \frac{1}{8\pi G_N} \left( \frac{V''}{V} \right)
  \,,
\end{eqnarray}
where a prime denotes derivatives with respect to $\phi$. The
slow-roll parameters are automatically suppressed at high energies
\cite{MWBH}, even for potentials that might otherwise be
considered too steep to drive inflation \cite{steep}.

To lowest order in the slow-roll approximation it is remarkably
easy to calculate the primordial perturbation spectra expected to
be generated by an inflaton on the brane. Quantum fluctuations of
a light scalar field ($|\eta|\ll1$) in four-dimensional de Sitter
spacetime ($\epsilon\ll1$) give the standard result for the power
spectrum of inflaton fluctuations on the Hubble scale ($k=aH$)
\begin{equation}
 \label{Pphi}
{\cal P}_{\delta\phi} \simeq \left( \frac{H}{2\pi} \right)^2 \,.
\end{equation}
This determines the dimensionless curvature perturbation on
uniform-density hypersurfaces \cite{MWBH}
\begin{equation}
 \label{Pzeta}
{\cal P}_\zeta \simeq \left( \frac{H^2}{2\pi\dot\phi} \right)^2
\,.
\end{equation}
This is conserved on large (super-Hubble) scales for adiabatic density
perturbations, even in brane-world gravity \cite{LMSW}, 
simply as a consequence of local energy conservation
\cite{WMLL,conserved}. Thus the standard expression (\ref{Pzeta})
gives the initial (primordial) density perturbations from
slow-roll inflation driven by an inflaton on the brane. Of course
the actual values for $H$ and $\dot\phi$ in terms of the scalar
field potential will differ due to the modified Friedmann equation
at high energies.

To determine the amplitude of gravitational waves we do need to
consider the five-dimensional metric perturbations. For a de
Sitter brane in AdS$_5$ the wave equation is separable. We find a
discrete zero-mode (as in the original Randall-Sundrum model) and
a continuum of massive modes in the four-dimensional effective
theory with $m^2>9H^2/4$ \cite{GarrigaSasaki}. These massive modes
are not excited by the de Sitter expansion -- they remain in their
vacuum state -- and hence we need only consider the spectrum of
tensor perturbations on super-Hubble scales from zero-mode vacuum
fluctuations \cite{LMW,Rubakov}
\begin{equation}
{\cal P}_T = F(V/\sigma) 64\pi G \left( \frac{H}{2\pi} \right)^2
\,,
\end{equation}
where $F(V/\sigma)$ represents the enhancement with respect to the
standard four-dimensional result. At low-energies we have $F\to1$
but at high energies we find $F\simeq 3V^2/4\sigma^2 \gg 1$.
Nonetheless one can verify that the standard consistency relation
between observables \cite{Lidsey}
\begin{equation}
\frac{{\cal P}_T}{{\cal P}_\zeta} \simeq -2n_T \,,
\end{equation}
still holds, where $n_T\equiv d\ln{\cal P}_T/d\ln k$ is the
spectral tilt of the tensor power spectrum. It is quite unexpected
that the non-trivial 5D modifications to the expected power
spectra from slow-roll inflation leave the four-dimensional
consistency relation unaltered \cite{Lidsey,Mariam}.

\subsection{Quantum fluctuations on the brane}

The above calculation would be expected hold to lowest order in
the slow-roll approximation where one neglects the coupling
between scalar field perturbations and the metric perturbations.
On the other hand there is a small but finite coupling between
inflaton field fluctuations and the metric even in slow-roll
inflation. In 4D general relativity Mukhanov \cite{M88} and Sasaki
\cite{S86} showed how to consistently couple the linear field
perturbations to the metric and hence one can derive small
corrections to the scalar field fluctuations (\ref{Pphi}) during
inflation.

One might expect something similar for an inflaton on the brane,
but there are important differences. Although deviations from 4D
gravity are small at low energies (indeed we have argued that for
long-wavelength perturbations the curvature perturbation $\zeta$
is conserved as in general relativity), at high energies the brane
perturbations are strongly coupled to bulk gravity. A perturbative
calculation, introducing the coupling at first-order in the
slow-roll parameters, shows that there may be 
a damping effect on the small
scale (sub-horizon) modes on a single brane at high energies
\cite{Mizuno}. This is not so surprising as the single brane in
AdS bulk is an open system and high frequency modes on the brane
are coupled to bulk gravitons that can escape to future null
infinity.
This highlights the need to treat the coupled brane-bulk system in
a consistent way in order to define the initial vacuum state on an
inflating brane. This remains another unsolved problem in brane
cosmology.

Remarkably little work seems to have been done to investigate the
quantum field theory of a boundary (brane) field coupled to a
higher-dimensional (bulk) field. Recently George \cite{George}
considered a linear oscillator in flat spacetime linearly coupled
to another linear oscillator on its boundary. Above a critical
coupling one finds an instability corresponding to a discrete
tachyonic bound state. This has been extended \cite{Mennim}, 
to the case of a Minkowski boundary in
an AdS bulk (i.e., a Randall-Sundrum geometry). Again we found a
critical coupling above which there exists an unstable bound
state. But even for weak coupling one finds a small
imaginary part in the frequency of resonant modes 
(quasi-normal modes \cite{Sanjeev}) of the coupled
brane-bulk oscillators, which indicates the slow decay of
oscillations on the brane. This is an example of an effect which
would not be seen in a dimensionally-reduced low-energy effective
theory which included only the zero-mode of the bulk field.

\section{Conclusions}

In summary, the brane-world has offered a novel perspective on
many of the unsolved problems in cosmology including the number of
spatial dimensions, the initial singularity and the cosmological
constant problem.

There remain many unsolved problems even in the simplest case of a
co-dimension one brane. There is no known solution for an isolated
black hole on the brane. Indeed it has been conjectured that there
is no static black hole solution on the brane \cite{Emparan}. Only a few special
cases are known for anisotropic brane cosmologies \cite{Fabbri}. Even linear
perturbations about FRW branes are restricted to special cases or
numerical solutions as the wave equation in the bulk is not in
general separable in a Gaussian normal coordinate system.

Some of these problems are techinical difficulties simply due to the extra
complication of an inhomogeneous extra dimension, but some
problems are also more fundamental. If we are seeking to describe a
single brane in an infinite extra dimension that the brane alone
does not form a closed system and we must specify initial data in
the bulk to determine the subsequent evolution.

A century after Einstein introduced physicists to four-dimensional
spacetime, brane-worlds offer a big new higher-dimensional
playground for relativists to explore. Cosmological models provide
one area to study. If we can understand more about gravity in
higher-dimensions then cosmology should also provide an
opportunity to test these exotic ideas against observational data.


\begin{theacknowledgments}
  I am grateful to the organisers for their hospitality in Oviedo
  and I thank my collaborators for the many collaborations upon
  which the work described in this article is based.
  This work is supported in part by PPARC grant PPA/G/S/2002/00576.
\end{theacknowledgments}

\end{document}